# The Covariant Electromagnetic Casimir Effect for Real Conducting Cylindrical Shells

**H. Razmi** [(1)] and **N. Fadaei** [(2)]

Department of Physics, The University of Qom, Qom 37185-359, I. R. Iran.
(1) razmi@qom.ac.ir & razmiha@hotmail.com  (2) fadaei.n@gmail.com

**Abstract**

Using covariant quantization of the electromagnetic field, the Casimir force per unit area experienced by a long conducting cylindrical shell, under both Dirichlet and Neumann boundary conditions, is calculated. The renormalization procedure is based on the plasma cut-off frequency for real conductors. The real case of a gold (silver) cylindrical shell is considered and the corresponding electromagnetic Casimir pressure is computed. It is discussed that the Dirichlet and Neumann problems should be considered separately without adding their corresponding results.

**Keywords:** The Casimir Effect; Canonical Quantization; Green Function

**PACS** 03.70.+k, 42.50.Lc

**Introduction**

After the simple geometry of two flat neutral conducting plates firstly introduced by Casimir [1], many theoretical and experimental researches have been conducted regarding calculation of Casimir force in other geometrical set ups [2]. Calculation of the Casimir effect in cylindrical geometry, in addition to its theoretical importance, has valuable applications in new technological field of studies (e.g. the role of the Casimir effect in nanotubes [3-6]). The Casimir effect in cylindrical geometry has been studied by a number of researchers. One of the first important computations of the Casimir stress on a perfectly conducting cylindrical shell can be found in [7]; the result is an attractive stress proportional to the inverse square radius of the cylinder. The vacuum energies of scalar fields under Dirichlet and Neumann boundary conditions, with opposite signs, on an infinite cylindrical surface have been computed in [8]. The electromagnetic Casimir energy for an infinite solid cylinder made of a material with specified dielectric and magnetic properties has been calculated in [9].

An exact versus semiclassical result for the Casimir interaction between two perfectly conducting, infinite, concentric cylinders can be found in [10]. The Casimir interaction energy between two perfectly conducting concentric cylinders and the force between slightly eccentric cylinders have been computed in [11]. Authors of [12-14], have calculated the exact Casimir interaction energy/force between two perfectly conducting, very long, eccentric cylindrical shells. An exact expression for the Casimir force between a plate and a cylinder can be found in [15]. Although each of these researches has its special technique (e.g. Green function or mode by mode summation method) of computing the Casimir energy/force, all of them deal with non-covariant formulation of the electromagnetic field. Indeed, in the already known calculations of the electromagnetic Casimir effect, the quantum vacuum fluctuations corresponding to the non-covariant quantities $\vec{E}$ and $\vec{B}$ (the electric and magnetic fields separately) are considered. As we know, the origin of the Casimir effect refers to the virtual particles (photons) of the quantum vacuum state; are these quantum field theoretical "virtual" particles appeared in a non-covariant quantum theory of the electric and magnetic fields? Obviously, these vacuum particles are quanta of the covariant electromagnetic field tensor $F^{\mu\nu}$ based on the four-vector potential $A^\mu$. Therefore, to have a better understanding of the quantum vacuum and the Casimir effect both conceptually and technically, we should work with the covariant quantization of the electromagnetic field. Here, we want to find the Casimir force/pressure experienced by a long real conducting cylindrical shell using covariant formulation of the electromagnetic field. Our calculation is based on a cut-off frequency regularization whose value is determined by the plasma frequency of the metal the cylindrical shell made of it. As we know, at low frequencies, all the metals/conductors have a real and frequency independent conductivity; but, at frequencies higher than the plasma frequency, the electromagnetic field- here the virtual photons causing the Casimir effect- "see" the conducting cylindrical shell transparent [16]. This means that for virtual photons with frequencies higher than the plasma frequency there is no boundary; in other words, there is no distinction between "in" and "out" of the shell and thus there is no Casimir effect. Only those photons whose frequencies are lower than the plasma frequency of the shell contribute to the Casimir effect. We restrict the upper bound of the integrals in the calculation of the Casimir force/pressure with plasma frequency cut-off. Although plasma frequency cut-off has been considered in the study of dielectrics (e.g. [17]), here we want to apply it to real conducting cylindrical shells.

A number of technical methods such as summation of modes [18], dimensional regularization [19], and Green function method [2] have been used to calculate the Casimir energy in different geometries. Among these well-known approaches, we use the Green function method based on a covariant formulation of the quantum theory of the electromagnetic field. The vacuum to vacuum expectation value of the covariant electromagnetic field tensor is found to be related to Feynman invariant delta function (time dependent Green function). After finding the appropriate Green functions for a long circular cylinder of radius $a$ with both Dirichlet and Neumann boundary conditions, the electromagnetic Casimir force per unit area is computed. The Dirichlet and Neumann problems are considered separately without adding their corresponding results.

**The electromagnetic Casimir effect: covariant formalism**

In non-covariant formulation of the electromagnetic field, only the transverse radiation field is quantized. Since the decomposition of the field into transverse and longitudinal components is frame-dependent, this clearly hides the Lorentz-invariance of the theory. To have a covariant theory, all four components of the four-vector potential $A^\mu(x) = (\phi, \vec{A})$, as the covariant dynamical variable under consideration, are quantized. The canonical stress tensor for the electromagnetic field[¶¶] is [20]

$$T^{\alpha\beta} = -g^{\alpha\mu} F_{\mu\lambda}(x) \partial^\beta A^\lambda(x) - g^{\alpha\beta} L_{EM} \quad (1)$$

where

$$L_{EM} = -\frac{1}{2} (\partial_\nu A_\mu(x))(\partial^\nu A^\mu(x)) \quad (2)$$

is the Fermi Lagrangian density[***] for the free electromagnetic field and

$$F^{\mu\nu}(x) = \partial^\mu A^\nu(x) - \partial^\nu A^\mu(x) \quad (3)$$

is the electromagnetic field tensor whose components are frame-dependent (non-covariant) quantities $\vec{E}$ and $\vec{B}$.

With the substitution of (2) and (3) in (1), it is found

$$T^{\alpha\beta} = -g^{\alpha\mu}(\partial_\mu A_\lambda(x) \partial^\beta A^\lambda(x) - \partial_\lambda A_\mu(x) \partial^\beta A^\lambda(x)) + \frac{1}{2} g^{\alpha\beta}(\partial_\nu A_\mu(x) \partial^\nu A^\mu(x)) \quad (4)$$

This can be written as

$$T^{\alpha\beta} = \lim_{x\to x'}\left[-g^{\alpha\mu}(\partial_\mu\partial'^\beta A_\lambda(x)A^\lambda(x')) - \partial_\lambda\partial'^\beta A_\mu(x)A^\lambda(x'))\right.$$

$$\left. + \frac{1}{2}g^{\alpha\beta}(\partial_\nu\partial'^\nu A_\mu(x)A^\mu(x'))\right] \quad (5)$$

Considering the operator form of the above relation and taking its vacuum to vacuum expectation value, it can be shown:

$$\langle 0|T^{\alpha\beta}|0\rangle = -i\hbar c \lim_{x\to x'}[-g^{\alpha\mu}(\partial_\mu\partial'^\beta g^\lambda_\lambda - \partial_\lambda\partial'^\beta g^\lambda_\mu) + \frac{1}{2}g^{\alpha\beta}(\partial_\nu\partial'^\nu g^\mu_\mu)]\Delta_F(x-x')$$

$$= -i\hbar c \lim_{x\to x'}[-g^{\alpha\mu}(4\partial_\mu\partial'^\beta - \partial_\mu\partial'^\beta) + \frac{1}{2}g^{\alpha\beta}(4\partial_\nu\partial'^\nu)]\Delta_F(x-x')$$

$$= i\hbar c \lim_{x\to x'}(3\partial^\alpha\partial'^\beta - 2g^{\alpha\beta}\partial_\nu\partial'^\nu)\Delta_F(x-x'), \quad (6)$$

in which we have used the following well-known relations:

$$\langle 0|T\{A^\mu(x)A^\nu(x')\}|0\rangle = i\hbar c D_F^{\mu\nu}(x-x'), \quad (7)$$

$$D_F^{\mu\nu}(x) = -g^{\mu\nu}\Delta_F(x) \quad (8)$$

$$\partial_\mu\partial^\mu\Delta_F(x-x') = -\delta^{(4)}(x-x') \quad (9)$$

$T\{A^\mu(x)A^\nu(x')\}$ is the time-ordered product of the field operators and $\Delta_F(x)$ is the famous Feynman delta-function (time dependent Green function) [20]. Thus

$$\langle 0|T^{\alpha\beta}|0\rangle = i\hbar c \lim_{x\to x'}(3\partial^\alpha\partial'^\beta - 2g^{\alpha\beta}\partial_\nu\partial'^\nu)G(x,x') \quad (10)$$

This means that to calculate the electromagnetic Casimir effect, it is enough to find the appropriate Green function corresponding to the geometry of the problem and then use the above relation for the vacuum to vacuum expectation value of the electromagnetic field stress tensor from which one can simply compute the desired Casimir force/pressure. The relation (10) is a general covariant formula which can be applied to different problems with different boundaries in flat (Minkowskian) space-time; this is because the metric tensor $g^{\mu\nu}$ with which we have worked is the Minkowskian metric tensor $g^{\mu\nu} = diag(1,-1,-1,-1)$. Of course, using quantum field theory in curved space-time, the method used here can be generalized for application in curved space-time geometries.

About the covariance of the formulation of the Casimir effect, we should mention that when a measurement is to be done, an experimentalist selects a specified (Lab.) frame and then experiences her/his aims. In other words, when one wants to find non-covariant quantities for a particular problem (e.g. force/pressure) with its special geometry and thus with its particular Green function, she/he ought to select an appropriate (special) frame and work with non-covariant quantities (e.g. the distance between two plates).

# Time dependent Green function (propagator) for a long conducting cylindrical shell

### A. Inside the shell

Working in cylindrical coordinates $(\rho, \varphi, z)$ and considering the special symmetry of the problem (an "infinitely" long cylindrical shell with circular basis of radius $a$), the Green function is found as:

$$G(x,x') = \frac{1}{(2\pi)^3} \sum_{m=-\infty}^{+\infty} \int_{-\infty}^{+\infty} dk \int_0^{\infty} d\omega \, e^{im(\varphi-\varphi')} e^{-i\omega(t-t')} e^{ik(z-z')} g(\rho, \rho') \qquad (11)$$

Application of the proper (Dirichlet and Neumann) boundary conditions leads to:

$$g_{in}^D(\rho, \rho') = (-\frac{\pi}{2ic}) \frac{J_m(\lambda \rho_<)}{J_m(\lambda a)} \left[ H_m^{(1)}(\lambda a) J_m(\lambda \rho_>) - J_m(\lambda a) H_m^{(1)}(\lambda \rho_>) \right] \qquad (12)$$

$$g_{in}^N(\rho, \rho') = (-\frac{\pi}{2ic}) \frac{J_m(\lambda \rho_<)}{J'_m(\lambda a)} \left[ H_m^{(1)'}(\lambda a) J_m(\lambda \rho_>) - J'_m(\lambda a) H_m^{(1)}(\lambda \rho_>) \right] \qquad (13)$$

where $\lambda = \sqrt{\frac{\omega^2}{c^2} - k^2}$, $J_m$ and $H_m^1$ are Bessel and Hankel functions of the first kind respectively, $\rho_<$ ($\rho_>$) is the smaller (larger) values of $\rho$ and $\rho'$, and

$$J'_m(\lambda a) = \frac{1}{\lambda} \frac{d}{d\rho} J_m(\lambda \rho)\Big|_{\rho=a}, \quad H_m^{(1)'}(\lambda a) = \frac{1}{\lambda} \frac{d}{d\rho} H_m^{(1)}(\lambda a)\Big|_{\rho=a}.$$

Using (11), (12) and (13), the proper Dirichlet and Neumann Green functions for the inside of the cylinder are found as the following:

$$G_{in}^D(x,x') = \frac{1}{(2\pi)^3} \left( -\frac{\pi}{2ic} \right)$$

$$\left( \sum_{m=-\infty}^{+\infty} \int_{-\infty}^{+\infty} dk \int_{-\infty}^{+\infty} d\omega \, e^{-i\omega(t-t')} e^{im(\varphi-\varphi')} e^{ik(z-z')} \frac{J_m(\lambda \rho_<)}{J_m(\lambda a)} \left[ H_m^{(1)}(\lambda a) J_m(\lambda \rho_>) - J_m(\lambda a) H_m^{(1)}(\lambda \rho_>) \right] \right)$$

$$\qquad (14)$$

$$G_{in}^N(x,x') = \frac{1}{(2\pi)^3}\left(-\frac{\pi}{2ic}\right)$$

$$\left(\sum_{m=-\infty}^{+\infty}\int_{-\infty}^{+\infty}dk\int_{-\infty}^{+\infty}d\omega\, e^{-i\omega(t-t')}e^{im(\varphi-\varphi')}e^{ik(z-z')}\frac{J_m(\lambda\rho_<)}{J'_m(\lambda a)}\left[H_m^{(1)\prime}(\lambda a)J_m(\lambda\rho_>)\right.\right.$$

$$\left.\left. - J'_m(\lambda a)H_m^{(1)}(\lambda\rho_>)\right]\right)$$

(15)

### B. Outside the shell

For the outside of the shell, the boundary conditions are: $g^D(\rho,\rho')\to 0$
($\frac{\partial}{\partial\rho}g^N(\rho,\rho')\to 0$) when $\rho\to a$ and $g(\rho,\rho')\to$ *free space Green function*
when $\rho\to\infty$. Therefore,

$$g_{out}^D(\rho,\rho') = \left(-\frac{\pi}{2ic}\right)\frac{H_m^{(1)}(\lambda\rho_>)}{H_m^{(1)}(\lambda a)}\left[J_m(\lambda a)H_m^{(1)}(\lambda\rho_<) - H_m^{(1)}(\lambda a)J_m(\lambda\rho_<)\right] \qquad (16)$$

$$G_{out}^D(x,x') = \frac{1}{(2\pi)^3}\left(-\frac{\pi}{2ic}\right)$$

$$\left(\sum_{m=-\infty}^{+\infty}\int_{-\infty}^{+\infty}dk\int_{-\infty}^{+\infty}d\omega\, e^{-i\omega(t-t')}e^{im(\varphi-\varphi')}e^{ik(z-z')}\frac{H_m^{(1)}(\lambda\rho_>)}{H_m^{(1)}(\lambda a)}\left[J_m(\lambda a)H_m^{(1)}(\lambda\rho_<)\right.\right.$$

$$\left.\left. - H_m^{(1)}(\lambda a)J_m(\lambda\rho_<)\right]\right)$$

(17)

$$g_{out}^N(\rho,\rho') = \left(-\frac{\pi}{2ic}\right)\frac{H_m^{(1)}(\lambda\rho_>)}{H_m^{(1)\prime}(\lambda a)}\left[J'_m(\lambda a)H_m^{(1)}(\lambda\rho_<) - H_m^{(1)\prime}(\lambda a)J_m(\lambda\rho_<)\right] \qquad (18)$$

$$G_{out}^N(x,x') = \frac{1}{(2\pi)^3}\left(-\frac{\pi}{2ic}\right)$$
$$\left(\sum_{m=-\infty}^{+\infty}\int_{-\infty}^{+\infty}dk\int_{-\infty}^{+\infty}d\omega\, e^{-i\omega(t-t')}e^{im(\varphi-\varphi')}e^{ik(z-z')}\frac{H_m^{(1)}(\lambda\rho_>)}{H_m^{(1)'}(\lambda a)}\left[J_m'(\lambda a)H_m^{(1)}(\lambda\rho_<)\right.\right.$$
$$\left.\left. - H_m^{(1)'}(\lambda a)J_m(\lambda\rho_<)\right]\right)$$

(19)

**The electromagnetic Casimir effect for a long real conducting cylindrical shell**

1. **Dirichlet problem**

The $\rho\rho$ component of the energy-momentum tensor (10) at the shell's surface is

$$\langle T^{\rho\rho}\rangle = i\hbar c\left[\frac{\partial}{\partial\rho}\frac{\partial}{\partial\rho'} - \frac{2}{\rho}\frac{\partial}{\partial\rho'} - \frac{2}{\rho\rho'}\frac{\partial}{\partial\varphi}\frac{\partial}{\partial\varphi'} + \frac{2}{c^2}\frac{\partial}{\partial t}\frac{\partial}{\partial t'}\right]G(x,x')\bigg|_{\substack{t\to t'\\ \varphi\to\varphi'\\ \rho,\rho'\to a}} \quad (20)$$

For the Dirichlet Green function, we simply arrive at the following result

$$\langle T^{\rho\rho}\rangle^D = i\hbar c\frac{\partial}{\partial\rho}\frac{\partial}{\partial\rho'}G^D(x,x')\bigg|_{\substack{t\to t'\\ \varphi\to\varphi'\\ \rho,\rho'\to a}} \quad (21)$$

By means of (14), it is found

$$\langle T^{\rho\rho}\rangle_{in}^D = \frac{i\hbar}{(2\pi)^3 a}\sum_{m=-\infty}^{+\infty}\int_{-\infty}^{+\infty}dk\int_{-\infty}^{+\infty}d\omega\,\lambda\frac{J_m'(\lambda a)}{J_m(\lambda a)}$$

(22)

Application of complex frequency rotation $(\omega\to i\omega)$, $(\lambda\to i\lambda; \lambda = \sqrt{\frac{\omega^2}{c^2}+k^2})$ and using the rectangular to polar coordinates transformation $\frac{1}{c}\int dk d\omega = \int_0^{2\pi}d\theta\int_0^\infty \lambda d\lambda$ lead to

$$\langle T^{\rho\rho}\rangle_{in}^D = \frac{-\hbar c}{(2\pi)^2 a}\sum_{m=-\infty}^{+\infty}\int_0^\infty d\lambda\,\lambda^2\frac{I_m'(\lambda a)}{I_m(\lambda a)}$$

(23)

in which $I_m$ is the modified Bessel function of the first kind.

Step by step repetition of all above works for the outside of the cylinder gives us

$$\langle T^{\rho\rho}\rangle^D_{out} = \frac{\hbar c}{(2\pi)^2 a} \sum_{m=-\infty}^{+\infty} \int_0^\infty d\lambda\, \lambda^2 \frac{K'_m(\lambda a)}{K_m(\lambda a)} \tag{24}$$

To calculate the force/pressure experienced by the shell, we need to find the following subtraction

$$\langle T^{\rho\rho}\rangle^D_{in} - \langle T^{\rho\rho}\rangle^D_{out} = \frac{-\hbar c}{(2\pi)^2 a} \sum_{m=-\infty}^{+\infty} \int_0^\infty d\lambda\, \lambda^2 \left( \frac{I'_m(\lambda a)}{I_m(\lambda a)} + \frac{K'_m(\lambda a)}{K_m(\lambda a)} \right) \tag{25}$$

Introducing the dimensionless variable $x = \lambda a$:

$$\langle T^{\rho\rho}\rangle^D_{in} - \langle T^{\rho\rho}\rangle^D_{out} = \frac{-\hbar c}{(2\pi)^2 a^4} \sum_{m=-\infty}^{+\infty} \int_0^\infty dx\, x^2 \left( \frac{I'_m(x)}{I_m(x)} + \frac{K'_m(x)}{K_m(x)} \right) \tag{26}$$

Except for $m=0$ term in (26), all the other terms are symmetrically equal to each other [21]; thus

$$\langle T^{\rho\rho}\rangle^D_{in} - \langle T^{\rho\rho}\rangle^D_{out} = \frac{-\hbar c}{4\pi^2 a^4} \left( \int_0^\infty x^2 dx\, \frac{d}{dx}\ln(I_0(x)K_0(x)) + 2\sum_{m=1}^\infty \int_0^\infty x^2 dx\, \frac{d}{dx}\ln(I_m(x)K_m(x)) \right) \tag{27}$$

The integral expressions in the relation (27) have infinite values; we have reached an irregular result that must be renormalized in some way. As we have explained in the introduction, we regularize the above relation with plasma frequency cut-off integration. Indeed, for frequencies higher than the plasma frequency corresponding to the material kind of the cylindrical shell, the surface of the shell behaves transparently and the virtual photons of the vacuum state do not "see" any boundary; this means for frequencies $\omega > \omega_p$, there is no difference between "in" and "out" and the upper bound of the above integral formula should be stopped at the cut-off value $x = \frac{\omega_p}{c} a$. This fact that a real material cannot constrain modes of the field with wavelengths much smaller (frequencies much higher) than a typical length (a cut-off frequency) scale has been studied and considered in [22] particularly for the Casimir scalar field energy in the geometry of plate(s) and sphere. Therefore, we can regularize the relation (27) as:

$$\left( \left\langle T^{\rho\rho} \right\rangle_{in}^{D} - \left\langle T^{\rho\rho} \right\rangle_{out}^{D} \right)_{regularized} = \frac{-\hbar c}{4\pi^2 a^4} \left( \int_0^{(x=\frac{\omega_p a}{c})} x^2 dx \frac{d}{dx} \ln(I_0(x)K_0(x)) \right.$$

$$\left. + 2\sum_{m=1}^{\infty} \int_0^{(x=\frac{\omega_p a}{c})} x^2 dx \frac{d}{dx} \ln(I_m(x)K_m(x)) \right) \quad (28)$$

For real situations depending on different conductors with different material kinds and thus with different plasma frequencies, $x_{cutoff}$ takes a variety of values, here we choose the following value

$$x_{cut\,off} \approx \frac{(1.37 \times 10^{16})(10^{-7})}{3 \times 10^8} \approx 4.56667 \quad (29)$$

This is related to the real case of a cylindrical shell made of gold with a plasma frequency of the order of $(1.37 \times 10^{16} \, rad/\sec)$ with a radius of 0.1 micrometer (100 nanometers) which is a distance scale compatible to the current world of experiments deal with the Casimir effect; it is a distance scale at which the Casimir effect is noticeable.

For the above value of $x_{cutoff}$, using **Mathematica** program, it is found

$$\int_0^{x \approx 4.56667} x^2 dx \frac{d}{dx} \ln(I_0(x)K_0(x)) = -10.6787 \quad (30)$$

For $m \geq 1$, it can be simply shown that the terms under sum in (28) decrease enough rapidly as $m$ increases. This is because for $x_{cutoff} < 1$ the integrals under sum on $m$ approach zero with increasing $m$. Indeed, the leading term in the relation (28) is the first integral (corresponding to the modified Bessel functions of zero order). Even for $x_{cutoff} \geq 1$ with a large but finite value, the sum on $m$ in (28) converges. Considering and comparing the following computation of the one hundred and one thousand terms of the series on $m$

$$\sum_{m=1}^{100} \int_0^{x \approx 4.56667} x^2 dx \frac{d}{dx} \ln(I_m(x)K_m(x)) == -43.3425 \quad (31)$$

$$\sum_{m=1}^{1000} \int_0^{x \approx 4.56667} x^2 dx \frac{d}{dx} \ln(I_m(x)K_m(x)) = -44.3152 \quad (32),$$

we can simply neglect the higher order terms.

The integral relation (28), keeping the terms up to $m = 1000$, is computed as

$$\left( \left\langle T^{\rho\rho} \right\rangle_{in}^{D} - \left\langle T^{\rho\rho} \right\rangle_{out}^{D} \right)_{regularized} \approx \frac{-\hbar c}{4\pi^2 a^4} (-10.6787 - 88.6304) = \frac{+99.3091 \hbar c}{4\pi^2 a^4} \quad (33)$$

To find the desired electromagnetic Casimir force/pressure, it is enough to integrate on the above expression as the following:

$$f_{EM}^{D} = \int_{0}^{2\pi} \left( \langle T^{\rho\rho} \rangle_{in}^{D} - \langle T^{\rho\rho} \rangle_{out}^{D} \right)_{regularized} a \, d\varphi \quad (34)$$

$$f_{EM}^{D} \approx \frac{+(15.8055)\hbar c}{a^{3}} \quad (35)$$

This is the electromagnetic Casimir force the unit area of the real conducting (gold) cylindrical shell (under Dirichlet boundary condition) experiences.

2. **Neumann problem**

For the Neumann Green function, using (20), the vacuum to vacuum expectation value of the $\rho\rho$ component of the electromagnetic energy-momentum tensor is found as:

$$\langle T^{\rho\rho} \rangle^{N} = i\hbar c \left( -\frac{2}{\rho\rho'} \frac{\partial}{\partial\varphi} \frac{\partial}{\partial\varphi'} - 2\frac{\partial}{\partial z}\frac{\partial}{\partial z'} + \frac{2}{c^2}\frac{\partial}{\partial t}\frac{\partial}{\partial t'} \right) G(x,x')^{N} \Big|_{\substack{t \to t' \\ \varphi \to \varphi' \\ z \to z' \\ \rho,\rho' \to a}}$$

(36)

For the inner part of the shell, according to (15), one can simply find

$$\langle T^{\rho\rho} \rangle_{in}^{N} = \frac{2i\hbar}{(2\pi)^{3}a} \sum_{m=-\infty}^{+\infty} \int_{-\infty}^{+\infty} dk \int_{-\infty}^{+\infty} d\omega \frac{1}{\lambda}\left(\frac{m^2}{a^2} + \lambda^2\right) \frac{J_m(\lambda a)}{J'_m(\lambda a)}$$

(37)

Applying complex frequency rotation $(\omega \to i\omega), (\lambda \to i\lambda; \lambda = \sqrt{\frac{\omega^2}{c^2} + k^2})$ and using the rectangular to polar coordinates transformation $\frac{1}{c}\int dk d\omega = \int_{0}^{2\pi} d\theta \int_{0}^{\infty} \lambda d\lambda$ lead to

$$\langle T^{\rho\rho} \rangle_{in}^{N} = \frac{-2\hbar c}{(2\pi)^{2}a} \sum_{m=-\infty}^{+\infty} \int_{0}^{\infty} d\lambda \left(\frac{m^2}{a^2} - \lambda^2\right) \frac{I_m(\lambda a)}{I'_m(\lambda a)}$$

(38).

For the outside of the cylinder, we arrive at the following expression:

$$\langle T^{\rho\rho} \rangle_{out}^{N} = \frac{2\hbar c}{(2\pi)^{2}a} \sum_{m=-\infty}^{+\infty} \int_{0}^{\infty} d\lambda \left(\frac{m^2}{a^2} - \lambda^2\right) \frac{K_m(\lambda a)}{K'_m(\lambda a)}$$

(39)

In terms of the dimensionless variable $x = \lambda a$:

$$\langle T^{\rho\rho}\rangle_{in}^{N} - \langle T^{\rho\rho}\rangle_{out}^{N} = \frac{-\hbar c}{2\pi^2 a^4} \sum_{m=-\infty}^{+\infty} \int_0^{\infty} dx\, (m^2 - x^2)\left(\frac{I_m(x)}{I'_m(x)} + \frac{K_m(x)}{K'_m(x)}\right)$$

(40)

Again, we have reached some infinite/irregular integrals. Based on the same reasoning for the Dirichlet problem, we use plasma frequency cut-off. Although, in comparison to the series for the Dirichlet problem, a coefficient $m^2$ exists here, the series still converges enough rapidly to neglect higher order terms. Using **Mathematica** program, for the real case of a gold cylindrical shell:

$$\int_0^{x\approx 4.56667} dx\, (-x^2)\left(\frac{I_0(x)}{I'_0(x)} + \frac{K_0(x)}{K'_0(x)}\right) = -12.0925$$

(41)

$$\sum_{m=1}^{100} \int_0^{x\approx 4.56667} dx\, (m^2 - x^2)\left(\frac{I_m(x)}{I'_m(x)} + \frac{K_m(x)}{K'_m(x)}\right) = 4.01046$$

(42)

$$\sum_{m=1}^{1000} \int_0^{x\approx 4.56667} dx\, (m^2 - x^2)\left(\frac{I_m(x)}{I'_m(x)} + \frac{K_m(x)}{K'_m(x)}\right) = 4.98219$$

(43)

Keeping the first 1000 terms of the series:

$$\left(\langle T^{\rho\rho}\rangle_{in}^{N} - \langle T^{\rho\rho}\rangle_{out}^{N}\right)_{regularized} \approx \frac{-\hbar c}{2\pi^2 a^4}(-12.0925 + 9.96438) = \frac{+(2.12812)\hbar c}{2\pi^2 a^4} \quad (44)$$

$$f_{EM}^{N} = \int_0^{2\pi} \left(\langle T^{\rho\rho}\rangle_{in}^{N} - \langle T^{\rho\rho}\rangle_{out}^{N}\right)_{regularized} a\, d\varphi \approx \frac{+(0.6774)\hbar c}{a^3}$$

(45)

This is the electromagnetic Casimir force the unit area of the real conducting (gold) cylindrical shell (under Neumann boundary condition) experiences.

**Conclusions**

We have studied the electromagnetic Casimir effect for a long real conducting cylindrical shell with a regularization method based on plasma frequency cut-off integration. The electromagnetic Casimir pressure experienced by the real case of a gold cylindrical shell has been computed. In two **Tables** at the end of this section, the electromagnetic Casimir force (per unit area) for the case of a cylindrical shell made of silver has been also computed. All the numerical integrals have been computed using **Mathematica** program software. Both the Dirichlet Casimir force and the Neumann one are repulsive. In the already known computations of the Casimir effect based on non-covariant formalisms, one needs to add the Dirichlet and Neumann Casimir forces/pressures to find the electromagnetic Casimir force/pressure. Here, we haven't added these two "different" results. This is because in non-covariant formalisms, the electric and magnetic fields contribution to the total energy of the electromagnetic field is considered separately and the fields are (often) divided to TE and TM modes with Dirichlet and Neumann boundary conditions; but here in the covariant approach, the Dirichlet and Neumann boundary value problems are two different/separate problems. In the covariant quantization of the electromagnetic field, one quantizes $A_\mu$ and doesn't work with the non-covariant objects $E_i$ and $B_i$. Although $E_i$ and $B_i$ have the main physical role of the electromagnetic fields in classical non-covariant arguments, we have quantized the covariant four potential $A_\mu$ (as a basis for the electromagnetic field tensor $F^{\mu\nu}$) as is usual in the standard canonical quantum theory of the electromagnetic field. Then, since the potential $A_\mu$ either satisfies Dirichlet or Neumann boundary conditions separately, we have led to this result that these two boundary conditions should be "disentangled".

As is clear from the data in the following tables, irrespective of the functional form of the forces, the numerical results directly depend on the cut-off values (material kinds); this is an unavoidable property of the Casimir energy for real materials [22].

**Table1** (Dirichlet boundary condition)

| Material | $\omega_p$ (rad/sec) | $a$ (meters) | $x_{cutoff}$ | $f_{EM}^D$ per unit $(\frac{\hbar c}{a^3})$ |
|---|---|---|---|---|
| Gold | $1.37 \times 10^{16}$ | $10^{-7}$ | 4.56667 | 15.8055 |
| Silver | $9.65 \times 10^{14}$ | $10^{-7}$ | 0.32167 | 0.0062 |

**Table2** (Neumann boundary condition)

| Material | $\omega_p$ (rad/sec) | $a$ (meters) | $x_{cutoff}$ | $f_{EM}^N$ per unit $(\frac{\hbar c}{a^3})$ |
|---|---|---|---|---|
| Gold | $1.37 \times 10^{16}$ | $10^{-7}$ | 4.56667 | 0.6774 |
| Silver | $9.65 \times 10^{14}$ | $10^{-7}$ | 0.32167 | 0.0286 |

¶¶¶ *Although the 0i<sup>th</sup> component of the electromagnetic stress tensor (1) doesn't initially identify with the linear momentum density (the Poynting vector) $(\vec{E} \times \vec{B})^i$,*

*because of the boundary conditions we are considering here and after necessary integral calculations, the additional terms in $T^{0i}$ are vanished finally and it acts as $(\vec{E} \times \vec{B})^i$.*

[***] *Fermi Lagrangian density is simply found from the well-known Lagrangian density $L = -\frac{1}{4} F_{\mu\nu} F^{\mu\nu}$ by choosing the covariant Lorentz gauge. It hasn't any difficulty in the introduction of the zero component of momentum (conjugate) field [20]. Although working in Lorentz gauge has other difficulties that can be removed with the formalism introduced by Gupta and Bleuler [23-24], our physical results here corresponding to the Casimir effect are independent of choosing any particular gauge.*